\documentstyle[11pt,aaspp4,tighten]{article}

\def\ksm{km~s$^{-1}$~Mpc$^{-1}$}

\def\ho{{\it H}$_{\rm 0}$}
\def\qo{{\it q}$_{\rm 0}$}
\def\h0{{\it H}$_{\rm 0}$}
\def\q0{{\it q}$_{\rm 0}$}

\def\p2c{{\it P}$_{\rm 20cm}$}

\def\ar{$\alpha_{\rm rad}$}

\def\lr{log($R$)}
\def\ebv{$E$($B$$-$$V$)}

\def\rr8{{\it R$_{8.4}$}}
\def\rrs{{\it R$^*$}}
\def\bk{$B$$-$$K$}

\lefthead{HALL, MARTINI, DEPOY, GATLEY} 
\righthead{BAL Quasars in the Near-IR}

\begin{document}

\title{The Optical/Near-IR Colors of Broad Absorption Line Quasars, Including the Candidate Radio-Loud BAL Quasar 1556$+$3517}

\author{Patrick B. Hall}
\affil{Steward Observatory, The University of Arizona, 933 N. Cherry Ave., Tucson, Arizona 85721 \\
Electronic Mail: pathall@as.arizona.edu}
\author{Paul Martini$^{1}$ and D.~L.~DePoy\footnote{Visiting Observers, Kitt Peak National Observatory, National Optical Astronomy Observatories, operated by the Association of Universities for Research in Astronomy under agreement with the National Science Foundation.}}
\affil{Astronomy Department, The Ohio State University, 174 West 18th Avenue, Columbus, OH 43210 
}
\and
\author{Ian Gatley}
\affil{Kitt Peak National Observatory, 950 N. Cherry Ave., Tucson AZ 85726\\
}

\begin{abstract}	\label{begin}

A candidate radio-loud broad absorption line quasar (RLBAL)
has been reported by Becker et~al. (1997).
We present JHK observations of this object
and three other radio-detected BALs
taken with the new 
Michigan-Dartmouth-MIT/Ohio~State/Aladdin IR Camera (MOSAIC)
on the KPNO 4-meter.  
The candidate RLBAL 1556$+$3517
has \bk=6.63, redder than all but one or two known $z$$>$1 quasars.
This strongly suggests the observed continuum of this quasar is reddened by
dust.
Even when this extreme reddening is taken into account 1556$+$3517 is still
probably radio-loud, although near-IR spectroscopy to measure its Balmer
decrement will be needed to verify this.
In addition,
since it is a flat-spectrum object, VLBI observations to determine the extent
(if any) to which beaming affects our estimate of its radio luminosity will be
needed before 1556$+$3517 can be unequivocally declared a radio-loud BAL.

We also use our data and data from the literature to show that optically 
selected BALs as a class have \bk\ colors consistent with the observed 
distribution for optically selected quasars as a whole.  
Thus there is currently no evidence that the tendency of optically selected
BALs to be preferentially radio-intermediate (\cite{fhi93}) is due to 
extinction artificially lowering estimated BAL optical luminosities.
However, as most quasar surveys, both radio and optical, would be
insensitive to a population of reddened radio-quiet BALs,
the existence of a large population of reddened BALs similar to 1556+3517 
cannot yet be ruled out.

\end{abstract}


\section{Introduction}	\label{introduce}

Broad absorption line quasars (BALs) exhibit absorption troughs blueward of
high-ionization (and occasionally low-ionization) emission lines in their 
rest-frame UV continua.
Since the definitive work of Stocke et~al. (1992, hereafter S92)
it has been held that no BALs are radio-loud quasars (RLQs),
defined as having either an absolute radio power 
$P$$\gtrsim$10$^{32.5}$~ergs~s$^{-1}$~Hz$^{-1}$
or a ratio of radio to optical luminosities $R$$\gtrsim$10.  
However, Francis, Hooper \& Impey (1993, hereafter F93) have shown that BALs 
are preferentially ``radio-intermediate" quasars (RIQs), with 0.2$<$\lr$<$1.
This may be because BALs are
radio-quiet quasars (RQQs) with higher than average optical extinction.
In principle this can be tested by measuring near-IR fluxes,
which are less sensitive to extinction and thus provide a better measurement
of the quasar luminosity.

Recently Becker et~al. (1997, hereafter B97) reported the discovery of a
candidate radio-loud BAL (RLBAL) 1556+3517 in the FIRST Bright Quasar Survey
(FBQS; Gregg et~al.~1996).
This object shows strong absorption below 2800~\AA\ rest
(B97, Fig.~1).~~Extinction from dust associated with the BAL gas
of $\sim$5$^{\rm m}$ at 2500~\AA\ (equivalent to \ebv$\sim$1;
\cite{sf92}, hereafter SF92) 
would be sufficient to make the quasar appear a RLQ even if it were
actually a RQQ, as mentioned by B97.
This is a large amount of extinction, but Egami et~al. (1996, hereafter E96)
find \ebv$\sim$0.41$-$0.84 for the unusual BALs 0059$-$2735 and Hawaii~167
based on their large Balmer decrements.
Again, this hypothesis can be tested with near-IR fluxes.

Since the confirmed existence of an RLBAL would have an impact on
theoretical models, 
and since near-infrared observations are useful for
understanding the 
overabundance of BALs among RIQs, we have obtained
such observations for several BALs and report the results in this Letter.
We adopt \ho=50~\ksm, \qo=0.1, $\Lambda$=0, consistent with B97.

\section{Observations and Data Reduction}	\label{observe}

We observed the first two published BALs from the FIRST survey (B97),
0840+3633 and 1556+3517, and two flat-radio-spectrum RQQ BALs (1212+1445
and 1246$-$0542) from Barvainis \& Lonsdale (1997, hereafter BV97). 
The observations were obtained on 20 \& 24 March 1997 at the 4-meter Mayall
telescope at Kitt Peak National Observatory with MOSAIC, the
Michigan-Dartmouth-MIT/Ohio State/Aladdin Infrared Camera. 
MOSAIC employs a $512 \times 1024$ Aladdin (InSb) array as detector and has
both imaging and spectroscopic capabilities.  
For these observations we used the $f$/7.6 camera which results in a plate 
scale of 0\farcs18 pixel$^{-1}$.  
Typical seeing during these nights was 1$''$ and 2$''$ for
20 \& 24 March, respectively.  
We obtained $JHK$ photometry of these four targets by employing a five-position
dither pattern on the array in each filter.  
The median of the five images in each filter was used as a sky frame
and subtracted from each individual frame.
We then flatfielded the images with the difference of dome flats taken with
the lights on and the lights off.  
Standard stars were observed and reduced in the same manner.  
Measurements were made using a 2\farcs5 synthesized circular beam.  
The mean of each set of five measurements is presented in Table 1, along with 
literature data on 0059$-$2735 (\cite{haz87}) and Hawaii 167 (\cite{cow94}).  
Our data on 1246$-$0542 agree well with Hyland \& Allen (1982), 
although the object appears $\sim$0.1 mag brighter in epoch 1997.23.

\section{Analysis and Discussion}	\label{discuss}

The colors of our objects, corrected for galactic extinction, 
are given in Table 1.  These corrections were determined using \ebv\ from
Burstein \& Heiles (1978, 1982) and $R_{\lambda}$ from Mihalas \& Binney (1981).
The colors are also shown in Fig. 1, which consists of three \bk\ histograms.
The bottom histogram (dashed) is predominantly optically selected
$z$$>$1 quasars from Francis (1996) and Srianand \& Kembhavi (1997).
The middle histogram consists of our six objects (solid)
and the six other $z$$>$1 BALs with \bk\ values in the literature (dotted).
1556+3517 is one of the reddest quasars known, even redder than Hawaii~167.
A two-sided Kolmogorov-Smirnov (K-S) test using the {\sc statistics} package
in {\sc stsdas} under {\sc iraf}\footnote{The Image Reduction and Analysis
Facility (IRAF) is distributed by the National Optical Astronomy
Observatories, which is operated by AURA, Inc., under contract to
the National Science Foundation.} gives $P$=64.65\% that the 
\bk\ distributions of our 12 BALs matches that of optically selected quasars.
However, if we exclude from the set of BALs Hawaii~167, 1556+3517, and
0840+3633, none of which were optically selected, we obtain $P$=99.95\% that
the \bk\ distributions of optically selected BALs and optically selected
quasars match.\footnote{This probability is correct but unusually high.
Resampling simulations show that a dataset of B-K colors of nine randomly 
chosen optically selected quasars would almost certainly give a lower but
still significant $P$.}
Thus there is no evidence for optically selected BALs as a class to have 
redder \bk\ colors than other optically selected quasars.
Therefore the observed tendency of optically selected BALs to be 
radio-intermediate (F93) does not seem to be 
due to extinction artificially lowering estimated BAL optical luminosities.
This extends to a longer wavelength baseline 
the finding of Weymann et~al. (1991) that the observed-frame optical continua
of high-ionization BALs and non-BALs are essentially identical.

However, the fact that neither of the two known examples of extremely red BALs
(Hawaii~167 and 1556+3517) were optically selected casts doubt on the
applicability of this conclusion to the entire BAL population, 
as distinct from the known optically selected BAL population.
A strong claim has been made by Webster et~al. (1995) that the color
distribution of radio-selected quasars extends to much redder colors
than that of optically-selected quasars.
To illustrate this, 
in the top histogram of Fig. 1 we plot the \bk\ distribution for a subsample
of radio-selected quasars from the Parkes 0.5~Jy survey (\cite{web95,dri97}),
including 23 objects with $z$$>$1 and 11 objects of unknown redshift.
The \bk\ distribution of radio-selected quasars
spans a range of \bk\ colors that includes those of 1556+3517 and Hawaii~167.
A two-sided K-S test gives 
$P$=5.11\% that the \bk\ distributions of BALS and radio-selected quasars match,
but $P$=0.83\% that the \bk\ distributions of optically selected BALs
and radio-selected quasars match.

This suggests that optically selected BALs may have a \bk\ distribution 
indistinguishable from that of optically selected quasars simply because BALs 
have been almost exclusively optically selected to date.
Whatever produces the bluer average \bk\ colors of optically-selected quasars 
(selection effect or lack of a red spectral component seen in RLQs)
would then naturally affect the colors of optically selected BALs as well.
This does not change our conclusion that extinction is unlikely to be
responsible for the observed preferential radio-moderateness of optically
selected BALs,
but it should be kept in mind that the observed distribution of 
BAL \bk\ colors may not represent the true distribution.
%
%
%
Radio surveys such as the FBQS (\cite{gre96}) which are sensitive to 
RQQs and RIQs 
to fairly high $z$ may help remedy this situation, 
since such surveys can in 
principle select BALs regardless of their optical/IR colors.

There are three obvious explanations for the extreme redness of 1556+3517:
dust, a beamed red synchrotron component, and/or contamination from its 
host galaxy.
The latter possibility can be almost completely ruled out, as it would require
a galaxy at $z$=1.48 with $K$$\lesssim$15.  This is two magnitudes brighter
than the brightest radio galaxies at $z$$\sim$1.5 (\cite{mcc93}).
%
The possibility of a beamed red synchrotron component (\cite{sr96}) in this
flat-spectrum source cannot be ruled out with photometry alone.
However, Francis et~al. (1997) show that for the Parkes radio-selected quasars
the equivalent width of Mg~{\sc ii} does not anticorrelate with continuum
redness, as expected if a red synchrotron component were present.
Thus there is good reason to believe that a red synchrotron component is
not present in most red quasars.
To determine this for certain for 1556+3517 specifically will require
measurement of its near-IR polarization (which should be low if such a 
component is absent) or of its Balmer decrement (which should be high if
dust is instead the reddening agent).

Assuming that the extreme redness of 1556+3517 is due to dust,
is its RLQ status due just to extinction?
We can test this by calculating $R$, the ratio of radio to optical luminosity.
Two specific definitions are in use:
$R^*$, the ratio of rest-frame 5~GHz to 2500~\AA\ flux densities (S92),
and $R_{8.4}$, the ratio of rest-frame 8.4~GHz to $B$ flux densities 
(\cite{hoo95}, hereafter H95).

To calculate the radio flux densities, we ignore the likelihood of variability
since we have only single-epoch data.
To change our conclusions, however, would require extreme radio variability 
of a factor $\gtrsim$3, unusual for quasars (\cite{haa92}).
We also assume isotropic radio emission, a point we shall return to later.
B97 use radio spectral index \ar=$-$0.1 based on 1.4 and 5~GHz detections of 1556$+$3517.
We use \ar=$-$0.1 for consistency except for 1212$+$1445 and 1246$-$0542,
for which multiwavelength radio data is available (BV97).  
We calculate \ar=$-$0.259$\pm$0.009 for 1212$+$1445 and
\ar=+0.097$\pm$0.021 for 1246$-$0542.
Flux densities and luminosities at 1400, 5000 and 8400 MHz are given in Table 2.

To calculate the true optical/UV flux densities, we again ignore variability
because of our lack of multi-epoch data.
We incorporate the differential $k$-correction between BALs and non-BALs
from S92, which is $\leq$0\fm05 except for 1246$-$0542 (0\fm25) and 
Hawaii~167 (0\fm3).
For Hawaii~167 and 1556+3517, because of the anomalously red \bk\ colors
we assume that the observed $B$ magnitudes do not reflect their intrinsic
$B$ luminosities, due to dust reddening and extinction.  
We want to find the $observed$ (dereddened) 
$B$ magnitudes the objects would have in the absence of dust.
We assume that without dust the objects would have observed \bk=3.
This is a plausible unreddened value for $z$$>$1 quasars
(Fig. 1; cf. \cite{web95,fra96,sk97}),
and just slightly bluer than the \bk\ of 0059$-$2735 and 0840+3633 
(cf. Table 1), two of the other three known ``Iron low-ionization BALs'' 
(the third is Mrk 231, \cite{smi95}). 
%
%
We determine the rest-frame \ebv\ necessary to produce the observed \bk\ 
colors assuming the dust is at the quasar $z$, which gives a conservative \ebv.
We use a SMC extinction law, in accordance with SF92 who found that LMC or MW 
reddening laws which have a 2200~\AA\ bump could not explain the observed 
reddening of low-ionization BALs like Hawaii~167 and 1556+3517.
We find \ebv=0.23 for Hawaii~167 and \ebv=0.63 for 1556+3517.
The \ebv\ for Hawaii~167 is $\sim$3.3
times lower than that derived by E96 from its Balmer decrement.
This is expected if, as they suggest, the quasar is highly extincted and
the observed rest-frame UV emission is actually from its host galaxy.
%
From these \ebv\ and Fig. 1 of SF92,
we find that 1556+3517 and Hawaii~167 have extinctions of 4\fm41 and 2\fm76 
in observed $B$ respectively, corresponding to 
dereddened $B$ magnitudes of 16.79 for 1556+3517 and 20.24 for Hawaii~167.

We then calculate \rrs\ using $S_{5000}$ and the dereddened $B$ magnitudes
in the formula log~$f_{2500\AA}$ = $-$22.62$-$0.4$B$ (S92).
We follow H95 in calculating absolute $B$ magnitudes (given in Table 1)
from the dereddened $B$ magnitudes using 
$k$-corrections for the composite 
Large Bright Quasar Survey spectrum of Francis et~al. (1991),
and then calculating \rr8\ using $L_{8.4}$ and 
the absolute rest-frame $B$-band luminosity $L_B$
(calculated assuming a zeropoint of 4260~Jy).
Values of \rrs\ and \rr8\ are given in Table 2.
0840+3633 and 1212+1445 appear to be RIQs by either definition of $R$.
1556+3517 is still a RLQ with log($R$)$\sim$1.43 even after our reddening 
correction.
However, if its true color was \bk=2 or 1 and we dereddened its $B$ magnitude 
appropriately, 1556+3517 would be a RIQ with \lr$\sim$1.03 or 0.73.
Thus a better estimate of the reddening for 1556+3517 is needed 
before it can be unequivocally declared a radio-loud quasar.
This could come from near-IR spectra to determine its Balmer decrement
or $L$ (3.4$\mu$m) or $M$ (5$\mu$m) data sampling rest-frame
$H$ and $K$, 
where extinction should be only 0\fm32 and 0\fm16 for \ebv=0.63.

The radio-loudness of 1556+3517
is in accord with its log($L$)$\sim$33.5 being above the usual divide
of 32.5 between RLQs and RQQs based on radio power alone (B97).
However, 1556$+$3517 is an unresolved flat-spectrum radio source, 
so the possibility remains that it is in fact a RQQ with a 
relativistically boosted jet pointed near our line of sight, and that we
have overestimated its radio luminosity by assuming isotropic radio emission.
This has been shown to be the case for several flat-spectrum RIQs
with \lr\ as high as 2.41 (\cite{fps96}).

\section{Conclusions}	\label{conclude}

We have shown that the candidate radio-loud BAL 1556+3517 has a redder
\bk\ color~than almost every other quasar known.  
Despite this, it is still probably a RLQ by all common~definitions,
unless it has a true (unreddened) observed color of \bk$<$2
or its red color is due to a beamed red synchrotron emission component.
Also, because it is a flat-spectrum
radio source, its radio luminosity may be affected by relativistic beaming.
Thus VLBI observations to determine the core dominance of 1556$+$3517
and near-IR spectroscopy to measure its Balmer decrement
will be needed to definitively determine if 1556$+$3517 is a radio-loud BAL.

We have also shown that 
at $z$$>$1, optically selected BALs as a class have a \bk\ color distribution
consistent with that of all optically selected quasars.
Thus there is currently no evidence that the tendency of optically selected
BALs to be preferentially radio-intermediate (F93) is due to extinction 
artificially lowering our estimate of BAL optical luminosities.
However, as most quasar surveys, both radio and optical, would not be
sensitive to a population of reddened radio-quiet BALs,
the existence of a large population of reddened BALs similar to 1556+3517
cannot yet be ruled out.

\acknowledgements

We thank E. Hooper for the composite LBQS $k$-corrections,
R. Srianand and P. Francis for providing data in electronic form, P. Osmer and
the referee for helpful comments, and P. Strittmatter for financial support.
This research has made use of data from the FIRST Survey, the NVSS,
and the NASA/IPAC Extragalactic Database (NED), operated by the Jet Propulsion
Laboratory, California Institute of Technology, under contract to NASA.

%



%
%

\onecolumn

\begin{figure}
\plotone{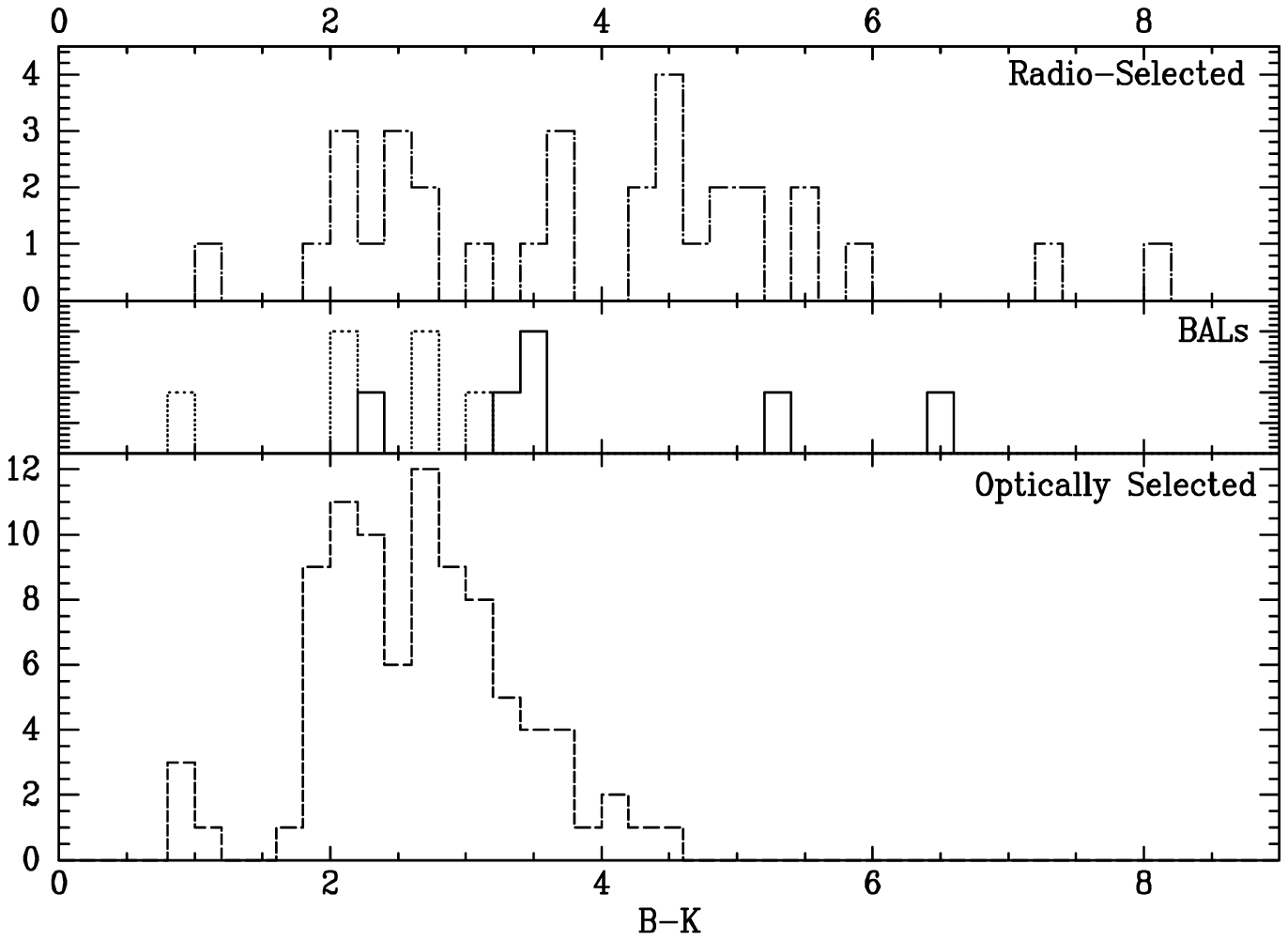}
\caption{
Histogram of quasar $B-K$ colors.
Bottom (Dashed):  $z$$>$1 quasars, primarily optically selected,
from Francis (1996) and Srianand \& Kembhavi (1997).
Middle: Data from this paper.
Solid: our BALs (see Table 1).
1556+3517 is the reddest of our objects, and Hawaii~167 the next reddest.
Dotted: BALs from the literature.
BALs were taken from Table~2 of Junkkarinen, Hewitt, \& Burbidge 1991,
and colors from the above two references.
In order of increasing \bk, these objects are 
2225$-$05, 1254+047, 0946+301, 2241+0016, 0004+147, and 1309$-$056.
Top (Dot-Dash): Radio-selected Parkes quasars from Webster et al. (1995),
including 23 objects with $z$$>$1 and 11 objects of unknown redshift.
}
\end{figure}

%
%
%
%
%
%
%


\begin{deluxetable}{lcccccccccc}
\scriptsize
\tablenum{1}
\tablewidth{530pt}
\tablecaption{BAL Quasar Optical-Infrared Properties}
\tablehead {
\colhead {ID} & 
\colhead {z} & \colhead {Galactic} & \colhead {B} & 
\colhead {J} & \colhead {H} & \colhead {K} &
\colhead {B-K} & 
\colhead {J-K} & \colhead {H-K} & \colhead {$M_B$} \\ 
\colhead {} & 
\colhead {} & \colhead {E(B-V)} & \colhead {} & 
\colhead {} & \colhead {} & \colhead {} &
\colhead {} & 
\colhead {} & \colhead {} & \colhead {} 
}
\startdata
%
0059$-$2735 
& 1.62 & 0.0203 & 18.2 
& 15.8 & 14.9 & 14.7
& 3.4 
& 1.1 & 0.2 & -$27.03$ \nl
Hawaii~167 
& 2.36 & 0.1443 & 23.0 
& 18.8 & 17.8 & 17.2
& 5.25 
& 1.5 & 0.6 & -$25.89$\tablenotemark{b} \nl
0840$+$3633 
& 1.22 & 0.0253 & 17.3 
& 14.87(0.07) & 14.10(0.10) & 13.92(0.06)
& 3.29 
& 0.94 & 0.18 & -$27.22$ \nl
1212$+$1445 
& 1.627 & 0.0312 & 17.94 
& 16.29(0.07) & 15.74(0.07) & 15.53(0.08)
& 2.29 
& 0.75 & 0.20 & -$27.29$ \nl
1246$-$0542 
& 2.236 & 0.0102 & 17.09 
& 14.96(0.06) & 14.34(0.04) & 13.61(0.05)
& 3.44 
& 1.34 & 0.72 & -$29.23$ \nl
1246$-$0542\tablenotemark{a} 
& & & 
& 15.20(0.04) & 14.43(0.02) & 13.71(0.02)  
& 3.34 
& 1.48 & 0.71 & \nl
1556$+$3517 
& 1.48 & 0.0143 & 21.2 
& 15.87(0.06) & 14.94(0.04) & 14.57(0.05)
& 6.57 
& 1.29 & 0.36 & -$28.18$\tablenotemark{b} \nl
\enddata
\tablecomments{
$H_{\rm 0}$=50, $q_{\rm 0}$=0.1, and $\alpha_{\rm opt}$=$-$1.0.
0059$-$2735 and Hawaii~167 magnitudes from Cowie et~al. (1994),
except 0059$-$2735 B magnitude from Warren, Hewett, \& Osmer (1991),
and 0059$-$2735 R magnitude from Becker et~al. (1997).
1212+1445 $B$ magnitude from Hooper et~al. (1995).
1246$-$0542 $B$ magnitude from Hewitt \& Burbidge (1993).
0840+3633 and 1556+3517 optical magnitudes from Becker et~al. (1997).
}
\tablenotetext{a}{Literature data from Hyland \& Allen 1982}
\tablenotetext{b}{Calculated using dereddened $B$ magnitudes
of 16.79 for 1556+3517 and 20.24 for Hawaii~167.  See text.}
\end{deluxetable}

\begin{deluxetable}{lcccccccccc}
\scriptsize
\tablenum{2}
\tablewidth{530pt}
\tablecaption{BAL Quasar Radio Properties}
\tablehead {
\colhead {ID} & 
\colhead {$S_{1400}$} & \colhead {$S_{5000}$} & \colhead {$S_{8400}$} &
\colhead {Log($L_{1400}$)} & \colhead {Log($L_{5000}$)} & \colhead {Log($L_{8400}$)} &
\colhead {Log($R^{*})$} & \colhead {Log($R_{8.4})$} \\
\colhead {} &
\multicolumn {3}{c}{(mJy)} &
\multicolumn {3}{c}{(ergs s$^{-1}$ Hz$^{-1}$)} &
\colhead {} & \colhead {}
}
\startdata
%
0059$-$2735 
& \nodata & $<0.36$\tablenotemark{c} & \nodata
& $<31.71$\tablenotemark{f} & $<31.66$ & $<31.64$\tablenotemark{f} & $<0.06$ & $<0.12$ \nl
Hawaii~167 
& $<2.5$\tablenotemark{a} & \nodata & \nodata
& $<32.85$ & $<32.79$\tablenotemark{h} & $<32.77$\tablenotemark{h} & $<1.58$ & $<1.71$ \nl
0840$+$3633 
& $\phm{<}1.6$\tablenotemark{b} & \nodata & \nodata
& $\phm{<}32.05$ & $\phm{<}32.00$\tablenotemark{h} & $\phm{<}31.97$\tablenotemark{h} & $\phm{<}0.38$ & $\phm{<}0.37$ \nl
1212$+$1445 
& $<0.60$ & 0.45(0.06) & 0.38(0.04)\tablenotemark{e}
& $<31.95$ & $\phm{<}31.82$ & $\phm{<}31.75$ & $\phm{<}0.12$ & $\phm{<}0.13$\tablenotemark{i} \nl
1246$-$0542 
& $<1.89$ & 0.52(0.05) & 0.53(0.06)
& $\phm{<}32.00$\tablenotemark{g} & $\phm{<}32.05$ & $\phm{<}32.06$ & $-0.49$ & $-0.34$ \nl
1556$+$3517 
& $\phm{<}30.6$\tablenotemark{b} & \phm{<}27.0\tablenotemark{d} & \nodata
& $\phm{<}33.50$ & $\phm{<}33.45$ & $\phm{<}33.43$\tablenotemark{h} & $\phm{<}1.41$ & $\phm{<}1.45$ \nl
\enddata
\tablecomments{
$H_{\rm 0}$=50, $q_{\rm 0}$=0.1, and $\alpha_{\rm rad}$=$-$0.1 
except 1212$+$1445 ($-$0.259$\pm$0.009) and 1246$-$0542 (+0.097$\pm$0.021).
$S$ values are monochromatic fluxes at the subscripted $observed$ frequency in MHz.
$L$ values are monochromatic luminosities at the subscripted $rest-frame$ frequency in MHz.
$R$ values are ratios of radio to optical luminosities as described in the text.
Fluxes for 1212$+$1445 and 1246$-$0542 from Barvainis \& Lonsdale 1997.
All upper limits are 3$\sigma$.
}
\tablenotetext{a}{Upper limit from the NVSS survey, Condon et~al. 1996}
\tablenotetext{b}{FIRST Survey, Becker, White, \& Helfand 1995}
\tablenotetext{c}{Stocke et~al.~1992}
\tablenotetext{d}{GB 5GHz survey, Becker et~al.~1991}
\tablenotetext{e}{Hooper et~al. (1995) give 0.45$\pm$0.07 mJy}
\tablenotetext{f}{Estimated from $S_{5000}$ upper limit}
\tablenotetext{g}{Estimated not from $S_{1400}$ upper limit but from higher frequency data in Barvainis \& Lonsdale 1997}
\tablenotetext{h}{Estimated from $S_{1400}$ detection or upper limit}
\tablenotetext{i}{Hooper et~al. (1995) give 0.31$\pm$0.32, calculated assuming $\alpha_{\rm rad}$=$-$0.3}
\end{deluxetable}




\begin{thebibliography}{}

\bibitem[Barvainis \& Lonsdale 1997]{bl97} \reference{}
Barvainis, R., \& Lonsdale, C. 1997, \aj, 113, 144 (BV97)

\bibitem[Becker, White, \& Helfand 1995]{bwh95} \reference{}
Becker, R.~H., White, R.~L., \& Helfand, D.~J. 1995, \apj, 450, 559

\bibitem[Becker et~al. 1997]{bec97} \reference{}
Becker, R.~H., Gregg, M.~D., Hook, I.~M., McMahon, R.~G., 
White, R.~L., \& Helfand, D.~J. 1997, \apjl, 479, L93 (B97)

\bibitem[Burstein \& Heiles 1978]{bh78} \reference{} Burstein, D., \&
    	Heiles, C.  1978, \apj, 225, 40

\bibitem[Burstein \& Heiles 1982]{bh82} \reference{} Burstein, D., \&
	Heiles, C.  1982, \aj, 87, 1165

\bibitem[Condon et~al. 1996]{con96} \reference{} Condon, J.~J., Cotton, W.~D.,
Greisen, E.~W., Yin, Q.~F., Perley, R.~A., and Broderick, J.~J.  1996, preprint

\bibitem[Cowie et~al. 1994]{cow94} \reference{}
Cowie, L.~L., Songaila, A., Hu, E.~M., Egami, E., Huang, J.-S., Pickles, A.~J,
Ridgway, S.~E., Wainscoat, R.~J., \& Weymann, R.~J. 1994, \apjl, 432, L83

\bibitem[Drinkwater et al. 1997]{dri97} \reference{}  Drinkwater, M.~J., 
Webster, R.~L., Francis, P.~J., Condon, J. J., Ellison, S. L., Jauncy, D. L.,
Lovell, J., Peterson, B.~A., and Savage, A.  1997, \mnras, 284, 85

\bibitem[Egami et~al. 1996]{ega96} \reference{} Egami, E., Iwamuro, F., 
Maihara, T., Oya, S., \& Cowie, L.~L. 1996, \aj, 112, 73 (E96)

\bibitem[Falcke, Patnaik, \& Sherwood 1996]{fps96} \reference{}
Falcke, H., Patnaik, A.~R., \& Sherwood, W.  1996, \apjl, 473, L13

\bibitem[Francis 1996]{fra96} \reference{}
Francis, P.~J. 1996, Pub. Astr. Soc. Australia, 13, 212

\bibitem[Francis et~al. 1997]{fra97} \reference{} 
Francis, P.~J., Webster, R.~L., Masci, F.~J., Drinkwater, M.~J., and
Peterson, B.~A.  1997, in ``IAU Colloquium 159: Emission Lines In Active 
Galaxies: New Methods and Techniques," eds. B.~M. Peterson, F.-Z. Cheng, and 
A.~S. Wilson, (San Francisco: ASP)

\bibitem[Francis et~al. 1991]{fra91} \reference{} 
Francis, P.~J., Hewett, P.~C., Foltz, C.~B., Chaffee, F.~C., Weymann, R.~J.,
\& Morris, S.~L. 1991, \apj, 373, 465

\bibitem[Francis, Hooper, \& Impey 1993]{fhi93} \reference{}
Francis, P.~J., Hooper, E.~J., \& Impey, C.~D. 1993, \aj, 106, 417 (F93)


\bibitem[Gregg et~al. 1996]{gre96} \reference{}
Gregg, M.~D., Becker, R.~H., White, R.~L., Helfand, D.~J., 
McMahon, R.~G., \& Hook, I.~M. 1996, \aj, 112, 407

\bibitem[Hazard et~al. 1987]{haz87} \reference{} Hazard, C., McMahon, R.~G., 
Webb, J.~K., \& Morton, D.~C.  1987, \apj, 323, 263

\bibitem[Hewitt \& Burbidge 1993]{hb93}\reference{}
Hewitt, A., \& Burbidge, G.  1993, \apjs, 87, 451

\bibitem[Hooper et~al. 1995]{hoo95} \reference{} Hooper, E.~J., Impey, C.~D., 
Foltz, C.~B., \& Hewett, P.~C.  1995, \apj, 445, 62 (H95)


\bibitem[Hughes, Aller, \& Aller 1992]{haa92} \reference{}
Hughes, P. A., Aller, H. D., and Aller, M. F.  1992, \apj, 396, 469

\bibitem[Hyland \& Allen 1982]{ha92} \reference{}
Hyland, A.~R., \& Allen, D.~A.  1982, \mnras, 199, 943

\bibitem[Junkkarinen, Hewitt, \& Burbidge 1991]{jun91} \reference{}
Junkkarinen, V., Hewitt, A., \& Burbidge G.  1991, \apjs, 77, 203

\bibitem[McCarthy 1993]{mcc93} \reference{}
McCarthy, P. J.  1993, \araa, 31, 639

\bibitem[Mihalas \& Binney 1981]{mb81} \reference{}
Mihalas, D. \& Binney, J.  1981, Galactic Astronomy, (New York: Freeman), 189

\bibitem[Serjeant \& Rawlings 1996]{sr96} \reference{}
Serjeant, S., and Rawlings, S.  1996, Nature, 379, 304

\bibitem[Smith et al. 1995]{smi95} \reference{}  Smith, P. S., Schmidt, G. D.,
Allen, R. G., and Angel, J. R. P.  1995, \apj, 444, 146

\bibitem[Sprayberry \& Foltz 1992]{sf92} \reference{}
Sprayberry, D., \& Foltz, C.~B. 1992, \apj, 390, 39 (SF92)

\bibitem[Srianand \& Kembhavi 1997]{sk97} \reference{}
Srianand, R., \& Kembhavi, A. 1997, \apj, 478, 70

\bibitem[Stocke et~al. 1992]{sto92} \reference{} Stocke, J.~T., Morris, S.~L.,
Weymann, R.~J. \& Foltz, C.~B.  1992, \apj, 396, 487 (S92)

\bibitem[Warren, Hewett, \& Osmer 1991]{who91} \reference{}
Warren, S.J., Hewett, P.C., \& Osmer, P.S. 1991, \apjs, 76, 1

\bibitem[Webster et~al. 1995]{web95} \reference{} Webster, R.~L., Francis, P.~J., Peterson, B.~A., Drinkwater, M.~J., \& Masci, F.~J.  1995, Nature, 375, 469

\bibitem[Weymann et~al. 1991]{wey91} \reference{} Weymann, R.~J., 
Morris, S.~L., Foltz, C.~B., \& Hewett, P.~C.  1991, \apj, 373, 23

\end{thebibliography}
\end{document}